\documentstyle[12pt]{article}
\author{Hans - J\"urgen Schmidt}
\title{Why do all the curvature invariants of a gravitational wave
 vanish ?
}
\date{}
\begin{document}
\maketitle

\bigskip

\centerline{
Universit\"at  Potsdam, Mathematisch-naturwiss. Fakult\"at}
\centerline{  Projektgruppe
Kosmologie}
\centerline{
      D-14415 POTSDAM, PF 601553, Am Neuen Palais 10, Germany}

\bigskip
\renewcommand{\baselinestretch}{1.4}
\begin{abstract}
We  prove  the  theorem valid  for  (Pseudo)-Riemannian  manifolds
$V_n$: "Let $x \in V_n$ be a fixed point of a  homothetic
motion  which  is not an isometry then all  curvature  invariants
vanish at $x$." and get the Corollary:  "All curvature invariants
of the plane wave metric $$ds \sp 2 \quad = \quad 2 \, du \, dv \, +
\, a\sp 2 (u) \, dw \sp 2 \,
 + \, b\sp 2 (u) \, dz \sp 2 $$ identically vanish."

Analysing the proof we see:  The fact that for definite signature Šflatness
can  be characterized by the vanishing of  a  curvature
invariant,  essentially  rests on the compactness of the rotation
group $SO(n)$.  For Lorentz signature,  however, one has the
 non-compact Lorentz group $SO(3,1)$ instead of it.

A  further and independent proof of the corollary uses the  fact,
that the Geroch limit does not lead to a Hausdorff topology,  so a
sequence  of gravitational waves can converge to the flat
 space-time, even if each element of the sequence is the same pp-wave.
\end{abstract}

\bigskip
AMS number: 53 B 30 Lorentz metrics, indefinite metrics

PACS number: 0430 Gravitational waves: theory
{\large
\bigskip

\section{ Introduction }

The   energy   of   the  gravitational   field   (especially   of
gravitational waves)
within  General Relativity was subject of controversies from  the
very  beginning,   see  [1]  and  the  cited  literature.  Global
considerations  - e.g.    by   considering   the   far-field   of
asymptotically  flat  space-times - soonly led  to  satisfactory
answers.  Local  considerations  became fruitful if a  system  of
reference  is  prescribed  (e.g.  by choosing a  time-like  vector
field).  If, however, no system of reference is preferred then it
is not a priori clear whether one can constructively  distinguish
flat space-time from a gravitational wave. This is connected with
the  generally  known  fact,  that for a pp-wave  (see  [2])  all
curvature  invariants  vanish  - but on the other  hand:  in  the
absence   of  matter  or  reference  systems   - only   curvature
invariants are locally constructively  measurable.

It  is the aim of this essay to explain the topological origin of
this strange property.

\bigskip

\section{ Preliminaries }
\setcounter{equation}{0}
  Let $V_n$ be a  $C \sp{\infty}$-(Pseudo)-Riemannian manifold
of arbitrary signature with dimension  $n>1$.
The metric and the Riemann tensor have components $g_{ij}$ and
$R_{ijlm}$  resp.  The  covariant derivative  with
respect  to  the coordinate $x \sp m $ is denoted by
";$m$" and is performed with the Christoffel affinity
$ \Gamma \sp i _{lm} $. We define

{\bf  Definition:   }  $I$  is  called  a  generalized  curvature
invariant of order $k$ if it is a scalar with dependence
$$ I = I(g_{ij},  \,  R_{ijlm}, \dots , R_{ijlm;i_1 \dots \, i_k}
).$$

By specialization we get the usual

{\bf  Definition:   }  $I$  is  called  a  curvature
invariant of order $k$ if it is a  generalized  curvature
invariant  of  order  $k$ which depends continuously on  all  its Šarguments.
The domain of dependence is requested to contain  the
flat space, and $ I(g_{ij}, \, 0, \dots \, 0) \equiv 0 $.

Examples: Let
$$ I_0 = sign (\sum \sp n _{i,j,l,m=1} \vert R_{ijlm} \vert ) $$
$I_0$  is  a generalized curvature invariant of order 0,  but  it
fails to be a curvature invariant.  It holds:  $V_n$ is flat  iff
(= if and only if) $I_0 \equiv 0$. Let further
$ I_1 = R_{ijlm} R \sp{ijlm} $
which  is  a curvature invariant of order 0.  If the  metric  has
definite signature or if $n=2$ then it holds:   $V_n$ is flat  iff
 $I_1 \equiv 0$.

{\it Proof:} For definite signature $I_0 = sign(I_1)$; for $n=2$,
$I_1 \equiv 0$ implies $R \equiv 0 $, hence flatness. $ \Box $

For  all other cases,  however,  the vanishing of $I_1 $ does not
imply flatness. Moreover, there does not exist another curvature
invariant serving for this purpose:

{\bf  Proposition:} For dimension $n \ge 3$,  arbitrary order  $k$
and indefinite metric it holds:  To each curvature invariant  $I$
of order $k$ there exists a non-flat $V_n$ with $I \equiv 0$.

{\it Proof:} Let $n=3$. We use
\begin{equation}
ds \sp 2 \quad = \quad 2 \, du \, dv \quad \pm
\quad a\sp 2 (u) \, dw \sp 2
\end{equation}
with a positive non-linear function $a(u)$. The "$\pm$" covers the
two possible indefinite signatures for $n=3$. The Ricci tensor
is $R_{ij} = R \sp m _{\quad imj}$ and has ($u = x \sp 1$)
\begin{equation}
R_{11} \quad = \quad - \, \frac{1}{a} \cdot
 \frac{d\sp 2 a}{du\sp 2}
\end{equation}
and therefore,  eq.  (2.1) represents a non-flat metric.  Now let
$n>3$. We use the cartesian product of (2.1) with a flat space of
dimension  $n-3$  and arbitrary signature.  So we have  for  each
$n\ge 3$ and each indefinite signature an example of a non-flat
$V_n$.  It  remains  to  show that for all  these  examples,  all
curvature  invariants of order $k$ vanish.  It suffices to  prove
that  at the origin of the coordinate system,  because at all  other
points  it  can  be  shown by  translations  of  all  coordinates
accompanied by a redefinition of $a(u)$ to $a(u-u_0)$. Let $I$ be
a curvature invariant of order $k$.  Independent of the dimension
(i.e.,  how many flat spaces are multiplied to metric (2.1))  one
gets for the case considered here that
$$I   \,    =   \,   I(a\sp{(0)}(u),   a\sp{(1)}(u),   \dots ,  \,
a\sp{(k+2)}(u)) $$
where $a\sp{(0)}(u) = a(u)$, $ a\sp{(m+1)}(u)
 = \frac{d}{du} a\sp{(m)}(u)$, and
$$I( a\sp{(0)}(u),  0,  \dots , \,  0) \,  = \, 0.$$ (This is because
each $R_{ijlm;i_1 \dots \, i_p}$
 continuously depends on $
a\sp{(0)}(u)$, $ a\sp{(1)}(u)$, $   \dots ,  \,
a\sp{(p+2)}(u)$
 and  on nothing else;  and for $a = $ const.,  (2.1) represents  a
flat space.)

Now  we apply a coordinate transformation:  Let $\epsilon > 0$ be Šfixed, we
replace $u$ by $u \cdot \epsilon$ and $v$ by
$v/  \epsilon$.  This  represents a Lorentz boost  in  the  $u-v-
$plane.  Metric  (2.1)  remains form-invariant by this  rotation,
only  $a(u)$  has to be replaced by $a(u \cdot  \epsilon  )$.  At
$u=0$ we have
$$I   \,    =   \,   I(a\sp{(0)}(0),   a\sp{(1)}(0),   \dots , \,
a\sp{(k+2)}(0)) $$
which must be equal to
$$I_{\epsilon} \, = \, I(a\sp{(0)}(0), \epsilon \cdot
 a\sp{(1)}(0),  \dots , \, \epsilon \sp{k+2} \cdot a\sp{(k+2)}(0)) $$
because $I$ is a scalar.  By continuity and by the fact that flat
space belongs to the domain of dependence of $I$, we have
$  \lim  _{\epsilon \rightarrow 0 } \,  I_{\epsilon}  =  0$.  All
values  $I_{\epsilon}  $ with $\epsilon >  0$  coincide,  and  so
$I=0$. $ \Box $

\bigskip

\section{Gravitational waves
}
\setcounter{equation}{0}
A  pp-wave (plane-fronted gravitational wave with parallel  rays,
see [2]) is a solution of Einstein's vacuum equation $R_{ij} = 0$
possessing a non-vanishing covariantly constant null vector.  The
simplest  type  of pp-waves can be represented similar as  metric
(2.1)
\begin{equation}
ds \sp 2 \quad = \quad 2 \, du \, dv \, +
\, a\sp 2 (u) \, dw \sp 2 \, + \, b\sp 2 (u) \, dz \sp 2
\end{equation}
 where
\begin{equation}
b \, \cdot \, \frac{ d \sp 2 a}{d u \sp 2} +
a \, \cdot \, \frac{ d \sp 2 b}{d u \sp 2}
\quad = \quad 0
\end{equation}
Metric (3.1) represents flat space-time iff both $a$ and $b$  are
linear  functions.   Using the arguments of sct.  2 one sees that
all curvature invariants of metric (3.1) identically vanish. Here
we present a second proof of that statement which has
the  advantage to put the problem into a more general  framework.
It holds

{\bf Theorem:} Let $x \in V_n$ be a fixed point of a  homothetic
motion  which  is not an isometry then all  curvature  invariants
vanish at $x$.

{\it Proof:}  The  existence  of  a homethetic motion which is  not  an
isometry  means  that $V_n$ is selfsimilar.  Let  the  underlying
differentiable  manifold  be equipped with two  metrics  $g_{ij}$
and   $ \tilde g_{ij} = e \sp{2C}  g_{ij}$  where $C$ is  a
non-vanishing  constant.  The corresponding Riemannian manifolds  are
denoted  by $V_n$ and  $ \tilde V_n$ resp.  By assumption,  there
exists  an  isometry from  $ V_n$ to  $ \tilde V_n$  leaving  $x$
fixed. Let $I$ be a curvature invariant. $I$ can be represented as
continuous  function (which vanishes if all the arguments do)  of
finitely  many  of  the  elementary  invariants.  The  elementary
invariants are such products of factors $g \sp{ij}$ with  factors
of  type   $R_{ijlm;i_1 \dots \,  i_p}$ which lead to  a  scalar,
i.e.,  all indices are traced out.  Let $J$ be such an elementary
invariant. By construction we have $J(x) = e \sp{qC} J(x)$ with a
Šnon-vanishing  natural  $q$ (which depends on the type  of  $J$).
Therefore, $J(x) = 0$. $\Box$

{\bf   Corollary:}  All  curvature  invariants  of  metric  (3.1)
identically vanish.

Remark:  This  refers  not only to the 14 independent  elementary
invariants  of  order 0,  see [3] for a list  of  them,  but  for
arbitrary order.

{\it Proof:}  We have to show that for each point $x$,  there exists  a
homothetic  motion with fixed point $x$ which is not an isometry.
But  this  is  trivially  done  by  suitable  linear   coordinate
transformations of $v$, $w$, and $z$. $\Box$

\bigskip

\section{Topological properties
}
\setcounter{equation}{0}
Sometimes it is discussed that the properties of space-time which
can  be  locally and constructively (i.e.,  by rods  and  clocks)
measured  are not only the curvature invariants but primarily the
projections of the curvature tensor and its covariant derivatives
to  an  orthonormal tetrad (4-bein).  (The  continuity  presumption
expresses the fact that a small deformation of space-time  should
also  lead to a correspondingly small change of the result of the
measurement.) To prevent a preferred system of reference one  can
construct curvature invariants like
\begin{equation}
I_2 \quad = \quad inf \quad \sum _{i,j,l,m} \quad \vert R_{ijlm}
\vert
\end{equation}
where the infimum (minimum) is taken over all orthonormal tetrads.
 From the first glance one could believe that $I_2 \, \equiv \, 0$
iff  the space is flat.  But for indefinite signature this  would
contradict the proposition of sct.  2.  What is the reason ?  For
definite signature the
infimum is to be taken about the rotation group $ SO(4) $
(or $ O(4) $ if one allows orientation-reversing  systems);  this
group is compact.  One knows: A positive continuous function over
a compactum possesses a positive infimum.  So, if one of the
$ R_{ijlm}$ differs from zero, then  $I_2 > 0$ at that point.
For Lorentz signature,  however, the infimum is to be taken about
the non-compact Lorentz group $SO(3,1)$ and so $ R_{ijlm} \ne 0$
does not imply $I_2 \ne 0$.

Another  topological  argument (which underlies our  sct.  2)  is
connected  with the Geroch limit of space-times [4],  we use  the
version of [5]. Theorem 3.1 of the first paper of ref. [5] reads:
(1)  For  local  Riemannian manifolds  with  definite  signature,
Geroch's  limit defines a Hausdorff topology.  (2) For indefinite
signature  this  topology  is not  even  $T_1$.  (A  topology  is
Hausdorff if each sequence possesses at most one limit, and it is
$T_1$ if  each constant sequence possesses at most one limit. The
main example is a sequence, where each element of the sequence is
the  same pp-wave,  and the sequence possesses two  limits:  flat
space and that pp-wave.)
  Here  the reason is:  Only for  definite  signature,
geodetic  $\epsilon$-balls  form a neighbourhood  basis  for  the
topology.      Š
\bigskip
{ \it References}}

{\large

[1] Einstein,  A.,  Sitz.-ber. Preuss. Akad. d. Wiss. Berlin 1914,
p. 1030; 1916 p. 688; 1918, p. 154.

[2] Stephani,  H., General Relativity, Cambridge University Press
1982, esp. sct. 15. 3. and references cited there.

[3] Harvey, A., Class. Quant. Grav. {\bf 7} (1990) 715; Lake, K.,
J. Math. Phys. {\bf 34} (1993) 5900 and cited references.

[4] Geroch, R., Commun. Math. Phys. {\bf 13} (1969) 180.

[5] Schmidt,  H.-J.,  J.  Math. Phys. {\bf 28} (1987) 1928;

 {\bf 29} (1988) 1264.}
\end{document}